\documentclass[aps,pra,twocolumn,showpacs,groupedaddress]{revtex4-1}
\usepackage{graphicx}  
\usepackage{dcolumn}   
\usepackage{bm}        
\usepackage{amssymb}   
\usepackage{amsmath}
\usepackage{braket}
\usepackage{enumitem}
\usepackage[mathscr]{euscript}
\usepackage{color}
\usepackage{epsfig,color}
\usepackage{hyperref}
\begin{document}
	
\title{Temporal Dynamics beyond the Exceptional Point in the Ikeda Map with Balanced Gain and Loss}
\author{Jyoti Prasad Deka}
\email{jyoti.deka@alumni.iitg.ac.in}
\affiliation{Department of Physics, Girijananda Chowdhury University, Guwahati-781017, Assam}
\author{Amarendra K. Sarma}
\email{aksarma@iitg.ac.in}
\affiliation{Department of Physics, IIT Guwahati, Guwahati-781039, Assam}
\date{\today}

\begin{abstract}
We investigate the temporal dynamics of the Ikeda Map with Balanced Gain and Loss and in the presence of feedback loops with saturation nonlinearity. From the bifurcation analysis, we find that the temporal evolution of optical power undergoes period quadrupling at the exceptional point ($\mathcal{EP}$) of the system and beyond that, chaotic dynamics emerge in the system and this has been further corroborated from the Largest Lyapunov Exponent ($\mathcal{LLE}$) of the model. For a closer inspection, we analyzed the parameter basin of the system, which further leads to our inference that the Ikeda Map with Balanced Gain and Loss exhibits the emergence of chaotic dynamics beyond the $\mathcal{EP}$. Furthermore, we find that the temporal dynamics beyond the $\mathcal{EP}$ regime leads to the onset of Extreme Events ($\mathcal{EE}$) in this system via attractor merging crisis. 
\end{abstract}
\pacs{XXXX}
\maketitle
Of the numerous models proposed to investigate the temporal instabilities of light, the Ikeda Map [1-2], Lorenz-Haken model [3-4] and the Lang-Kobayashi model [5] play an influential role in such endeavours. The Ikeda map is a discrete-time nonlinear system proposed to describe the temporal evolution of light propagating in a ring cavity filled with a dielectric nonlinear media. Chaotic dynamics emerge in the system as a consequence of the time-delayed feedback in the system. The Lorenz-Haken model is a nonlinear system of coupled differential equations which describes the analogy between a fluid close to the convection instability and a homogenously broadened single-mode gas laser close to the lasing threshold. And last of all, we have the Lang-Kobayashi Model, a nonlinear time-delayed system, which delineates the temporal dynamics of a semiconductor laser with external optical feedback. This model has been considered to be a historical milestone in research concerning chaotic dynamics in semiconductor lasers [6-10]. Over the previous three decades, there has been numerous reports on the nonlinear dynamics in semiconductor lasers both in theoretical and experimental domains and phenomena such as synchronization [11], stochastic resonance [12-13], quorum sensing [14], crowd synchrony [14], etc. have been reported in such systems. Furthermore, in the recent years with the advent of all-optical reservoir computing [15-19], nonlinear optical systems with time-delayed feedback are playing a vital role in the experimental realization of recurrent neural networks that are indispensable in such computational endeavours. Hence, it could be that optical systems have indeed facilitated the experimental realization of numerous exotic nonlinear phenomena that are observed in their mathematical modelling.

Now, Carl M. Bender and his student Stefan Boettcher reported that certain non-Hermitian Hamiltonians that are invariant under the joint operation of the parity ($\mathcal{P}$) and time-reversal ($\mathcal{T}$) operator possess a real eigenspectra [20-23]. Such Hamiltonians are defined as ($\mathcal{PT}$) Symmetric Hamiltonians. The presence of Exceptional Point ($\mathcal{EP}$) in the eigenspectra is an interesting feature of such Hamiltonians, which signifies a phase transition of the eigenspectra from real to imaginary or vice-versa, preceded by a coalescence of the eigenvalues. Several years later, Prof. D. N. Christodoulides and his research group at the University of Central Florida reported the novel idea that optics could provide the background for the experimental realization of the $\mathcal{PT}$-Symmetric quantum potentials [24]. And in 2012, R\"uter \textit{et al.} demonstrated the observation of $\mathcal{PT}$-Symmetry in evanescently coupled waveguide structure with balanced gain and loss [25]. Since then, $\mathcal{PT}$-Symmetry has been investigated in complex optical potentials [26], optomechanics [27-28], optical lattices [29], microring lasers [30], solitons [31-33], wireless power transfer [34], multilayered structures [35-36], many-body ultracold systems [37], Li\'enard oscillators [38] and so on.

\textit{Modelling} - In this work, we report on the exotic nonlinear phenomena that could be observed in the Ikeda Map with Balanced Gain and Loss (see Appendix for the schematic and the detailed mathematical modelling of the system). Previously, we investigated a similar model with Kerr nonlinearity and we reported  optical power saturation, blow-up of the $\mathcal{LLE}$ and intermittent surge in optical power [39]. Now, our focus is on the temporal dynamics in the Ikeda Map with Saturation Nonlinearity and with Balanced Gain and Loss in the two feedback loops of the system. In addition, we shall concentrate mostly on the regime beyond the $\mathcal{EP}$ of the system. The mathematical model of the system could be expressed as given below.
\begin{subequations}
	\begin{align}
	& E_{1,j+1}=ie^{i\psi \left( |E_{1,j}|^2 \right)} A + e^{i\left[\psi_{-} \left( |E_{2,j}|^2 \right) + \psi \left( |E_{1,j}|^2 \right) \right]} B \\
	& E_{4,j+1}=ie^{i \psi \left( |E_{4,j}|^2 \right)} C + e^{i\left[\psi_{+} \left( |E_{3,j}|^2 \right) + \psi \left( |E_{4,j}|^2 \right) \right]} D
	\end{align}
	\label{eq3}
\end{subequations}
where $A = e^{-\gamma/2}E_{in}/\sqrt{2}$, $B = e^{-\gamma} \left(E_{1,j}+iE_{4,j}\right)/2 $, $C = e^{\gamma/2}E^{'}_{in}/\sqrt{2}$ and $D = e^{\gamma}\left(iE_{1,j}+E_{4,j}\right)/2$. $\gamma$ is the gain/loss coefficient and saturation nonlinearity is incorporated into the model through the terms $\psi \left( |E_{i,j}|^2 \right) = \beta \left[1 + \eta \left( |E_{i,j}|^2 \right) \right]^{-1}$ and $\psi_{\pm} \left( |E_{i,j}|^2 \right) = \beta e^{\pm \gamma}\left[1 + \eta \left( |E_{i,j}|^2 \right) \right]^{-1}$, where $\beta$ is the coefficient and $\eta$ is the strength of saturation nonlinearity. $j$ is the iteration number and physically, one iteration step is characterized by the time taken by light to make a complete loop around the feedback loop in the two feedback loops.

In the absence of nonlinearity ($\eta = \beta = 0$), Eq. 1a and 1b could be rewritten in the matrix form as follows.
\begin{equation}
\begin{pmatrix} E_{1,j+1} \\ E_{4,j+1} \end{pmatrix}=
T
\begin{pmatrix} E_{1,j} \\ E_{4,j} \end{pmatrix} +
\frac{i}{\sqrt{2}} \begin{pmatrix} e^{-\gamma/2} E_{in} \\ e^{\gamma/2} E^{'}_{in} \end{pmatrix}
\end{equation}
where $T = \frac{1}{2} \begin{pmatrix} e^{-\gamma} & i e^{-\gamma} \\ i e^{\gamma} & e^{\gamma} \end{pmatrix}$ is the transfer matrix of the system. The eigenvalues of $T$ are $\lambda_{1,2} = \left ( cosh \gamma \pm \sqrt{cosh^2 \gamma -2} \right)/2$ and the exceptional point of this system could be evaluated to be $\gamma_{th}=cosh^{-1} \sqrt{2} = 0.88116$, where `$th$' stands for \textit{threshold}. This has been further elucidated in the plots in Fig. 1.
\begin{figure}
	\centering
	\includegraphics[height=4cm,width=9cm]{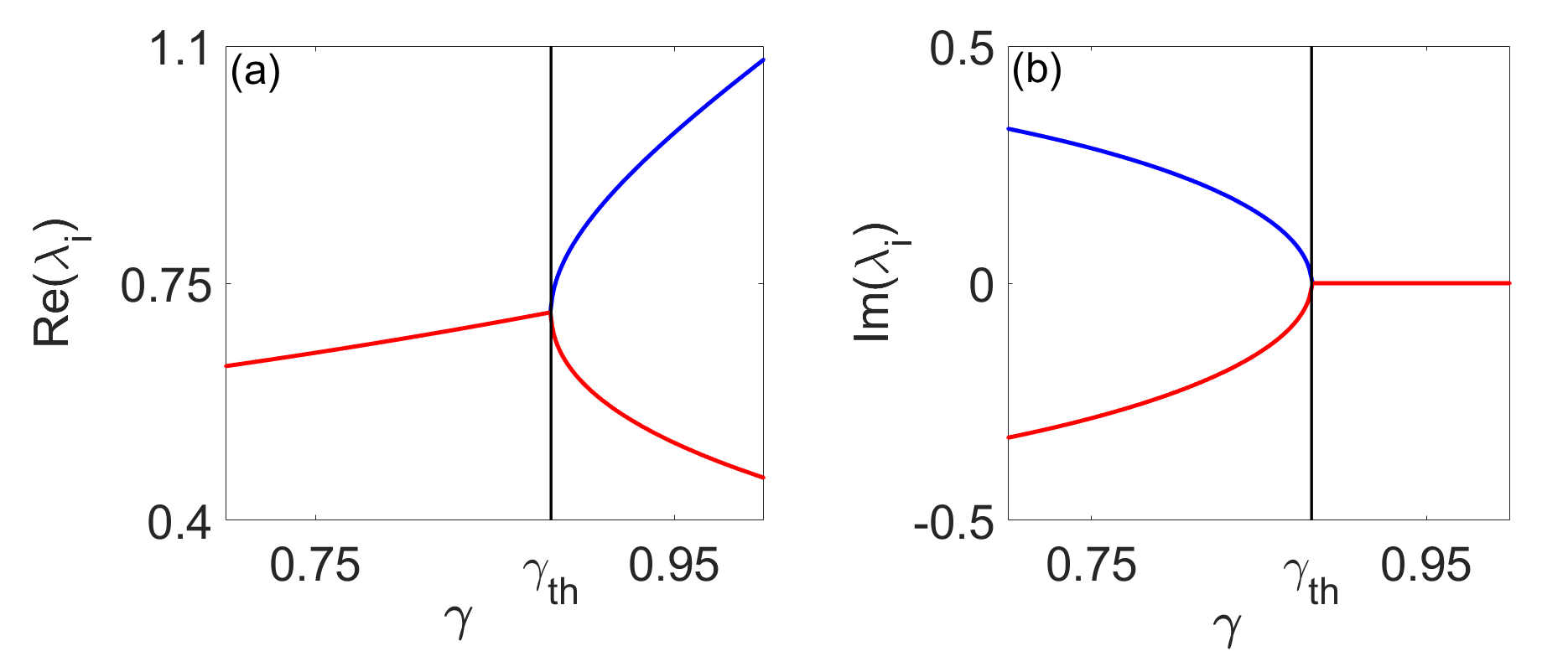}
	\caption{Eigenspectra of the Transfer Matrix $T$ for $\beta=\eta=0$. The \textbf{black-colored} vertical line is drawn at the $\mathcal{EP}$ of the system.}
\end{figure}

\textit{Onset of Chaotic Dynamics} - From the spectral analysis in Fig. 1, it could be clearly seen that at $\gamma_{th}=0.88116$, the real component of the eigenvalues merges into the x-axis, whereas the imaginary component bifurcates into two distinct divisions. We will now investigate the temporal dynamics of the system in the presence of saturation nonlinearity and for our analysis, we have chosen $\beta=1$ and $\eta=0.001$. In Fig. 2(a), we have presented the bifurcation analysis of the optical power in the amplified feedback loop  $P_4$ as a function of the gain/loss coefficient $\gamma$. For our analysis, we have considered up to 3500 iterations and we have considered 1500 iterations as transient and 2000 iterations as steady state. First of all, we see that as $\gamma$ is increased, there is tiny window between $\gamma=0.7$ and $\gamma=0.735$ where $P_4$ suddenly starts to display quasiperiodic oscillation and this could further be seen from the $\mathcal{LLE}$ plot in Fig. 2(b). In this region, the $\mathcal{LLE}$ is seen to be approximately zero and this characterizes the quasiperiodic dynamics exhibited by the system. From $0.735 \leq \gamma \leq \gamma_{th}$, the system exhibits period - 2 temporal evolution and at the $\mathcal{EP}$, this undergoes a phase transition from period - 2 to period - 4 temporal evolution and it further undergoes bifurcation into period - 8 dynamics at $\gamma=0.8945$ before descending into the chaotic regime. And hence, we can say that our system exhibits the period doubling cascade to chaos. Moreover, it must be noted that within the chaotic regime, there are regions where we observe period dynamics and from this, we can infer that our system exhibits intermittency in the parametric regime of $\gamma$. This has been further validated by the $\mathcal{LLE}$ of the system in Fig. 2(b). In the $0.735 \leq \gamma \leq \gamma_{th}$ regime, the $\mathcal{LLE}$ is a negative quantity which means that the period - 2 temporal evolution of optical power in the two two feedback loops is asymptotically stable. Beyond the $\mathcal{EP}$, the system is seen to exhibit stable period - 4 evolution. But such dynamics eventually descend into the chaotic regime as $\gamma$ is further increased. 
\begin{figure}
	\centering
	\includegraphics[height=4cm,width=9cm]{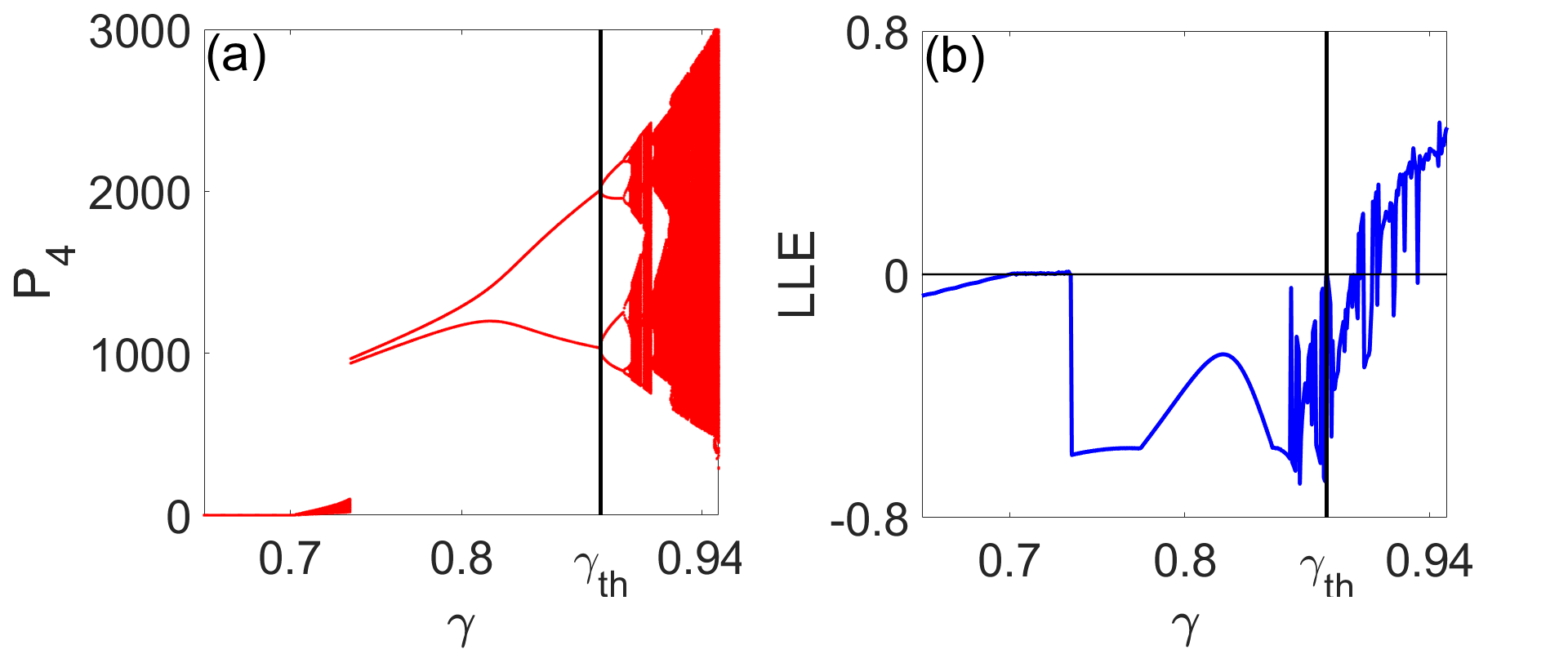}
	\caption{(a) Bifurcation Diagram and (b) $\mathcal{LLE}$ of the Ikeda Map with Balanced Gain and Loss for $E_{in}=E^{'}_{in}=0.65$.}
\end{figure}
\begin{figure}
	\centering
	\includegraphics[height=6cm,width=7cm]{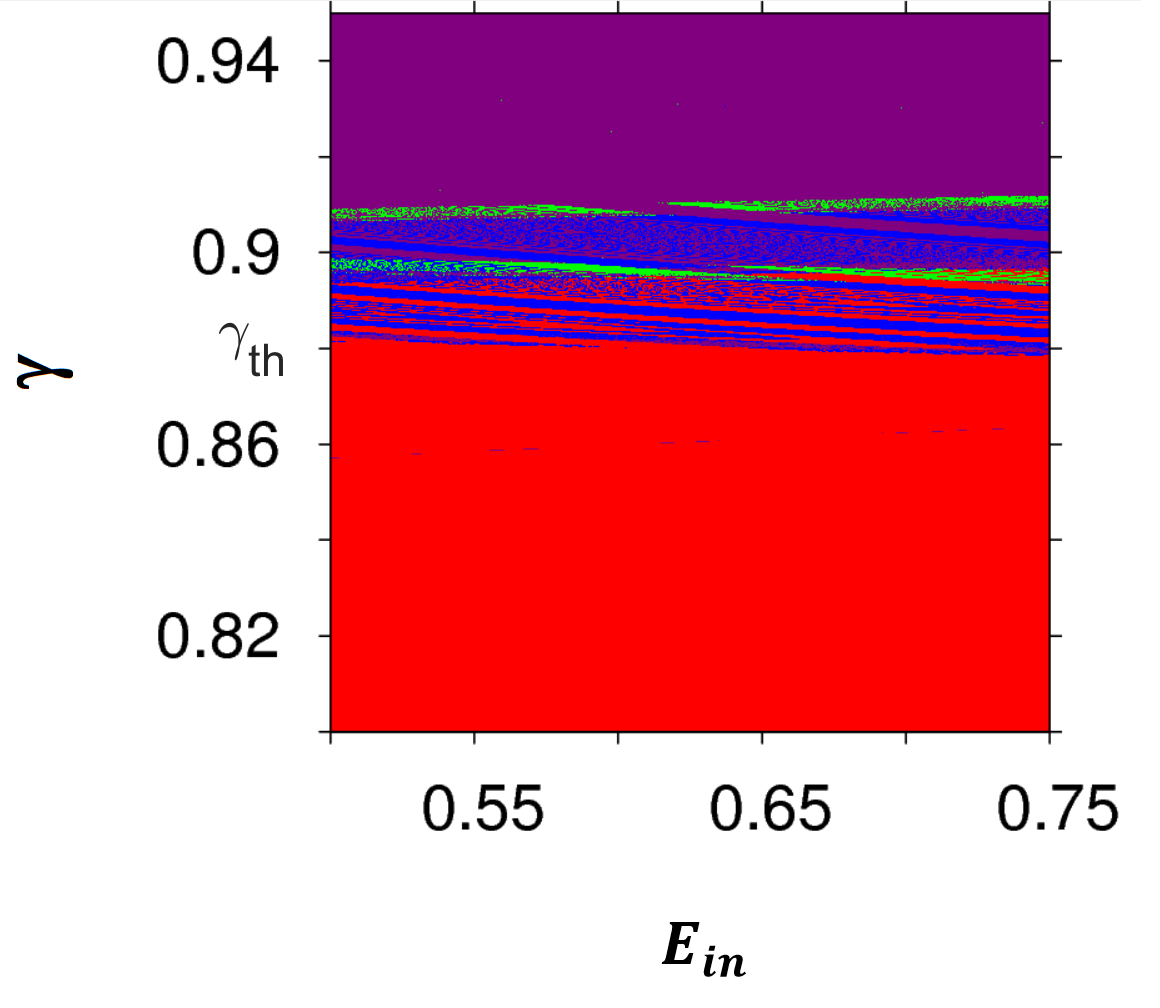}
	\caption{Parameter Basin for Periodic Temporal Evolution. \textbf{Red} - Period 2, \textbf{Blue} - Period 4, \textbf{Green} - Period 8 and \textbf{Purple} - Chaotic.}
\end{figure}

To analyze the system in-depth, we have presented the parameter-basin of our model as a function of $\gamma$ and the input field amplitude, taking into account $E_{in}=E^{'}_{in}$ in Fig. 3. This would give us additional leverage before we can conclude that our system exhibits chaotic dynamics beyond the $\mathcal{EP}$ for any choice of input field amplitudes. So, one of the first inference that we can have from Fig. 3 is that at the $\mathcal{EP}$ of the system, the optical power in the two feedback loops undergoes bifurcation from period - 2 to period - 4 temporal evolution and furthermore, there are regions in the parameter basin beyond the $\mathcal{EP}$ where we could observe period - 8 temporal evolution of optical power. But the purple region in the parameter basin which indicates chaos is predominant far beyond the $\mathcal{EP}$ and this, we believe, is sufficient evidence for us to conclude that our system portrays chaotic dynamics beyond the $\mathcal{EP}$ of the system.

\begin{figure}
	\centering
	\includegraphics[height=4cm,width=8cm]{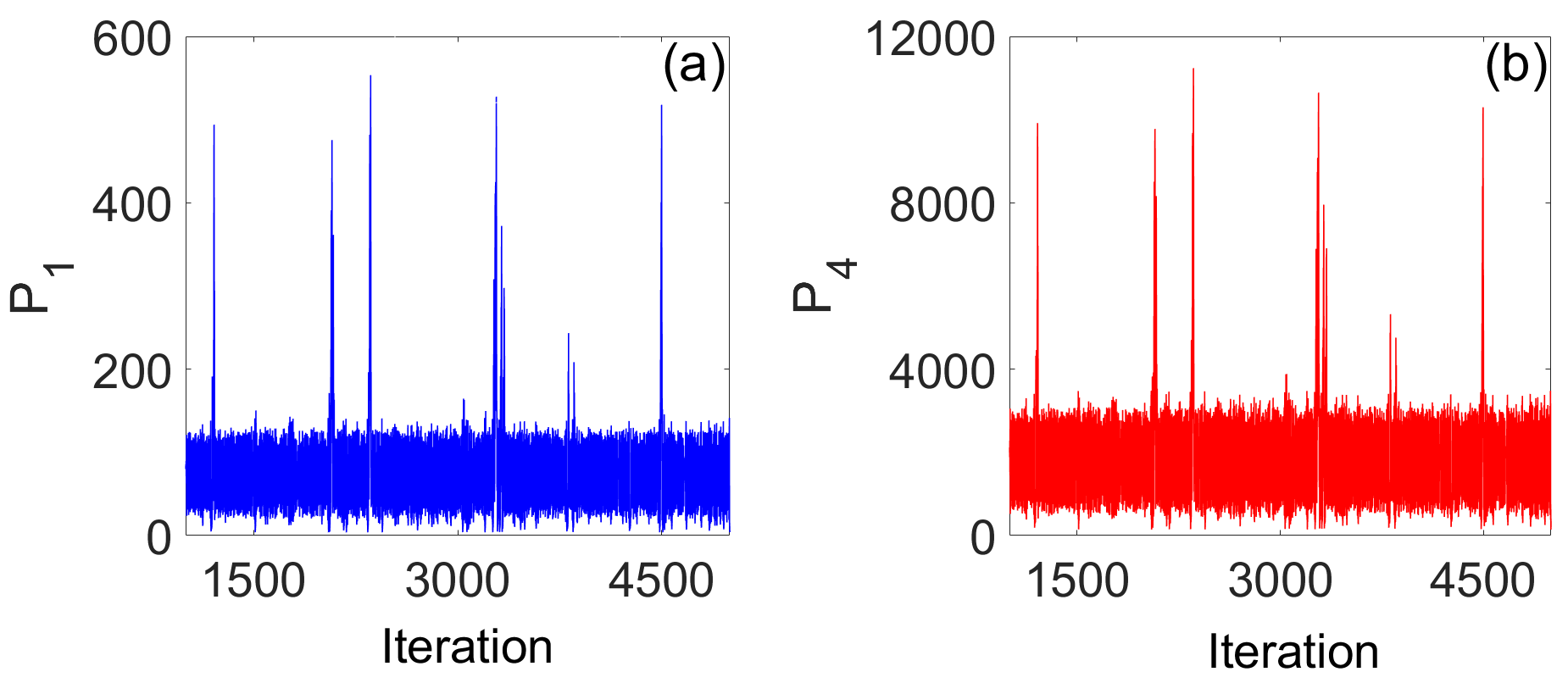}
	\caption{Temporal evolution of optical power depicting the emergence of $\mathcal{EE}$s for $\gamma=0.96$.}
\end{figure}

\textit{Emergence of Extreme Events} - Our investigation of the temporal dynamics further beyond the $\mathcal{EP}$  regime revealed the interesting emergence of Extreme Events ($\mathcal{EE}$s) in the system. In nonlinear optical systems, the emergence of such phenomenon could be attributed to the Benjamin-Feir (Modulational) Instability for certain choice of parameters and nonlinearity. $\mathcal{EE}$s have been reported in optical fibers [40-41], space plasmas [43], optically injected semiconductor laser [44], Li\'enard oscillators [45]  and so on. The temporal evolution of optical power in the two feedback loops of our system depicting the emergence of $\mathcal{EE}$s has been shown in Fig. 4. First of all, it should be noted that $\mathcal{EE}$s are observed in our model for $\gamma=0.96$, which is far beyond the $\mathcal{EP}$ of the system. $\mathcal{EE}$s in any system could be characterized by a histogram analysis of the time-series and this has been presented below in Fig. 5, where we used probability normalization of the data. It could very well be seen that the histogram depicts two sharp peaks for high optical power and it is infinitesimal in the extremely high optical power regime. This means that our system exhibits chaotic evolution of low optical power and intermittently it also exhibits an abrupt sharp surge in the optical power, which we can observe in Fig. 4. Even though the probability of the sharp surge in optical power is infinitesimally small, it shall still occur. Furthermore, our histogram analysis depicts a Poisson-type probability distribution for the time-series and as such, we can conclude that our system also exhibits $\mathcal{EE}$s of optical power beyond the $\mathcal{EP}$ of the system. 

\begin{figure}
	\centering
	\includegraphics[height=7cm,width=8cm]{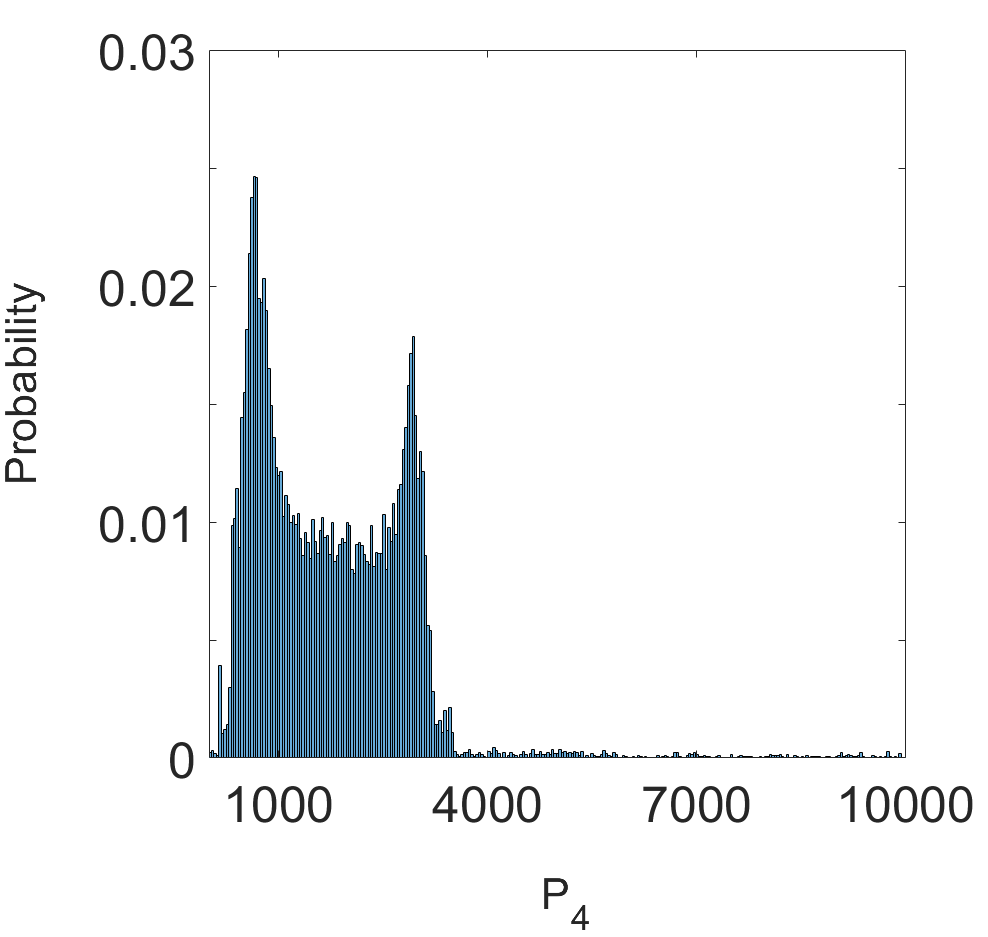}
	\caption{Probability histogram of the Temporal Evolution of $P_4$ for $\gamma=0.96$.}
\end{figure}

We would now like to investigate the temporal evolution of the field amplitude $E_4$ of the amplified feedback loop, so as to ascertain the cause for the onset of $\mathcal{EE}$s in the system. It could be seen from Fig. 6(a) that the temporal evolution of the imaginary component of $E_4$ ($y_4$) depicts two distinct regions clearly demarcated by the x-axis. These regions form the attracting sets of the time-series. But as $\gamma$ is increased to $\gamma=0.945$ in Fig. 6(b), it could be seen that the two attracting sets are tending to approach each other and for $\gamma=0.95$ in Fig. 6(c), the attracting
\begin{figure}
	\centering
	\includegraphics[height=8cm,width=8cm]{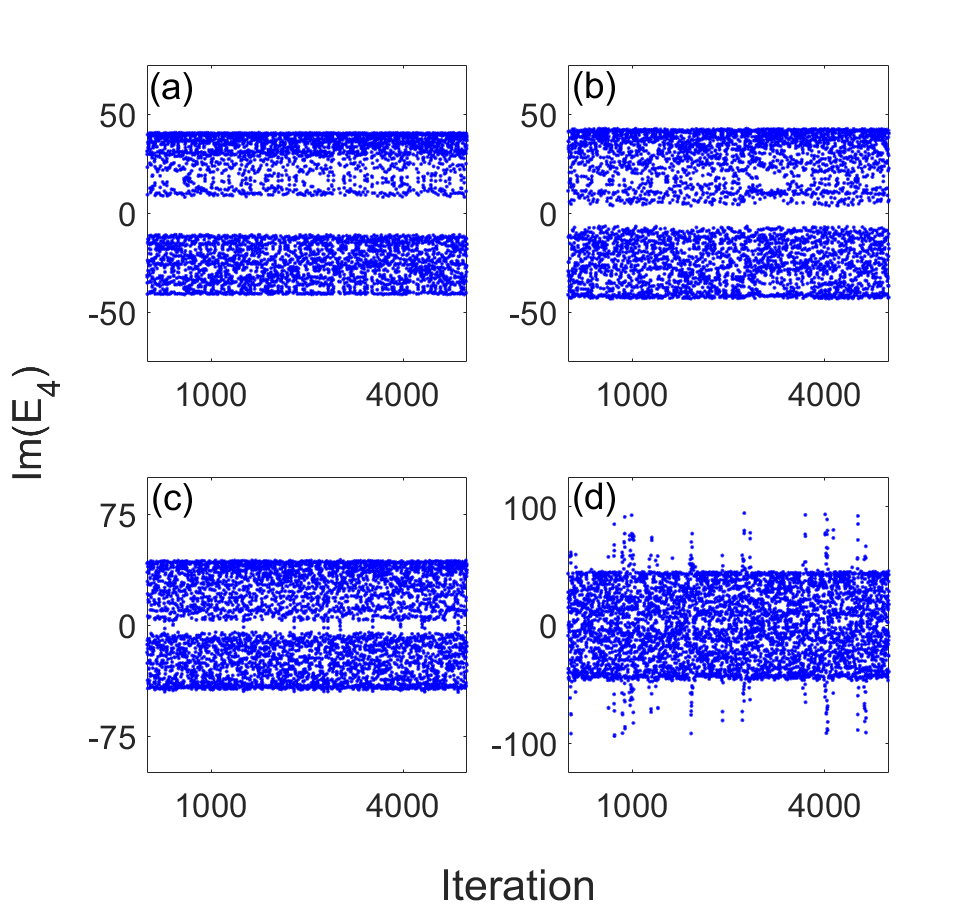}
	\caption{Temporal evolution of $Im(E_4)$ for (a) $\gamma=0.93$, (b) $\gamma=0.945$, (c) $\gamma=0.95$ and (d) $\gamma=0.96$.}
\end{figure}
sets could be seen to merge with each other. And when finally, the attracting sets merged into each other for $\gamma=0.96$ as shown in Fig. 6(d), we observe the emergence of $\mathcal{EE}$s in the system. Furthermore, in Fig. 7, we have plotted the $Re(E_4)$ vs. $Im(E_4)$ and the corresponding probability histogram for the two cases shown in Fig. 6(c) and Fig. 6(d). It could be seen in Fig. 7(a) and 7(b) that for $\gamma=0.953$, the $Re(E_4)$ vs. $Im(E_4)$ plot is showing the onset of the merge of the two attracting sets of the system and the probability histogram is depicting two peaks. For $\gamma=0.957$ on the other hand, the attracting sets could be seen to merge with each other and intermittent separation from the attractor takes place and this has been further corroborated by the probability distribution histogram in Fig. 7(d). Hence, from this, it could be concluded that our system exhibits the attractor merging crisis route to $\mathcal{EE}$s.  

\textit{Conclusion} - We have investigated the Ikeda Map with Balanced Gain and Loss beyond the $\mathcal{EP}$ regime and discussed the emergence of chaotic dynamics in the system. It has been shown that temporal evolution of optical power in the system undergoes a phase transition from stationary to quasiperiodic dynamics, which is then followed by a period doubling cascade to chaos. The discussion has further been corroborated by the $\mathcal{LLE}$ of the system. Our further analysis of the system beyond the $\mathcal{EP}$ led us to the emergence of $\mathcal{EE}$s, which occurred due to attractor merging crisis. Hence, it could be said that our analysis has thrown light into the generation of Chaotic Dynamics and $\mathcal{EE}$s in the Ikeda Map with Balanced Gain and Loss by tuning the gain/loss coefficient of the system externally and the same could be said for the experimental distinguishability of temporal dynamics in time-delayed feedback systems with saturation nonlinearity.
\begin{figure}
	\centering
	\includegraphics[height=8cm,width=8cm]{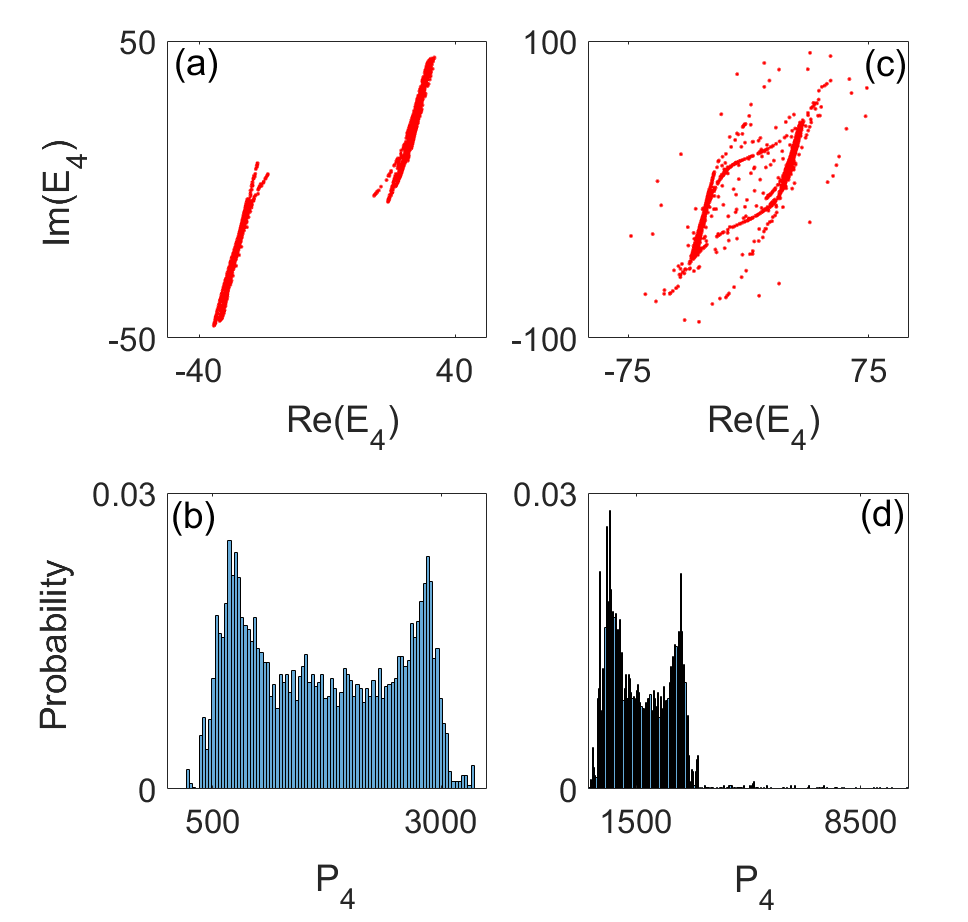}
	\caption{$Re(E_4)$ vs. $Im(E_4)$ and its corresponding probability histogram for (a, b) $\gamma=0.953$ and (c, d) $\gamma=0.957$.}
\end{figure}

\section*{Appendix}
The \textit{Ikeda Map with Balanced Gain and Loss} is a configuration of two input channels and two time-delayed nonlinear feedback loops interacting via a passive 50:50 directional coupler. The schematic of the model is presented in Fig. 8. 
\begin{figure}
	\centering
	\includegraphics[height=6cm,width=8cm]{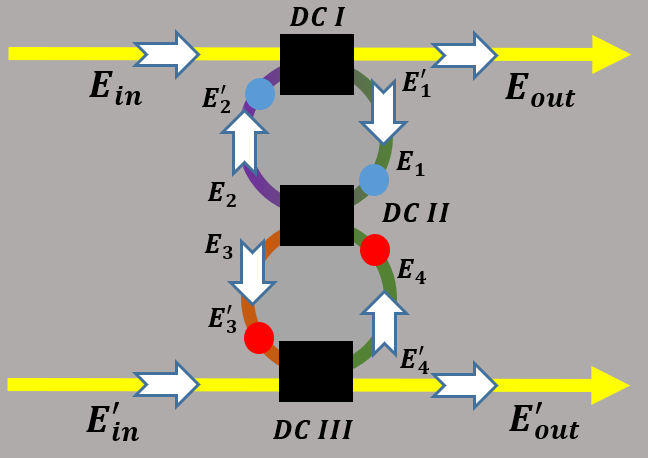}
	\caption{Schematic of the $\mathcal{PT}$-Symmetric Ikeda Map. The \textbf{yellow-colored} blocks are the passive input and output channels and the \textbf{black-colored} blocks, labelled as \textbf{\textit{DC I}}, \textbf{\textit{DC II}} and \textbf{\textit{DC III}}, are the 50:50 directional couplers. The amplifiers are colored \textbf{red} and the attenuators are colored \textbf{blue}. The feedback loops with saturation nonlinearity are colored green, purple and brownish.}
\end{figure}
The two feedback loops (colored green) transporting the field amplitudes $E_1$ and $E_2$ exhibit saturation nonlinearity of equal strength. Now, as the field amplitude $E^{'}_1$ propagates through the attenuated feedback loop from the output port of coupler region ‘\textbf{\textit{DC I}}’ to the attenuator, it suffers a linear and nonlinear phase shift $\phi_L$ and $\phi_{NL,1}$ and after it exits the attenuator, the phase-shifted field amplitude suffers loss as well in the blue-colored attenuator region. This could be then expressed in terms of the field amplitude $E_1$ as follows. 
\begin{equation}
E_1=e^{i(\phi_{NL,1}+\phi_L)} e^{-\gamma/2} E^{'}_1
\end{equation}
where $\phi_{NL,1}=\psi \left( |E_{1}|^2 \right)=\beta(1+\eta|E_1|^2)^{-1}$ is the nonlinear phase shift imparted to the field as it propagates through the feedback loop. In addition to this, the field amplitude also suffers loss by a factor of $e^{-\gamma/2}$. Here, ‘$\gamma$’ is the gain/loss parameter, ‘$\beta$’ is the coefficient and $\eta$ is the strength of the nonlinear phase shift and ‘$\phi_L$’ is the linear phase shift. In addition, we know
\begin{equation}
E^{'}_1=\frac{iE_{in}+E^{'}_2}{\sqrt{2}}
\end{equation}
where $E^{'}_2=e^{i\left(\phi_{NL,2}^{'}+\phi_L\right)}e^{-\gamma/2}E_2$ and $\phi_{NL,2}^{'}=\psi_{-} \left( |E_{2}|^2 \right)=\beta e^{-\gamma}(1+\eta|E_2|^2)^{-1}$. Using the transfer matrix of coupler region ‘\textbf{\textit{DC II}}’, $E_2$ and $E_3$ could be expressed in terms of $E_1$ and $E_4$ as follows.
\begin{equation}
\begin{pmatrix} E_2 \\ E_3 \end{pmatrix}=M \begin{pmatrix} E_1 \\ E_4 \end{pmatrix}
\end{equation}
where $M$ is the transfer matrix of the coupler region ‘\textbf{\textit{DC II}}’ and it is given by $M=\frac{1}{\sqrt{2}} \begin{pmatrix} 1 & i \\ i & 1 \end{pmatrix}$. Using Eq. 4 and Eq. 5 in Eq. 3, we have
\begin{equation}
E_1=e^{i\left(\phi_{NL,1}+\phi_L\right)} e^{-\gamma/2} \left(\frac{iE_{in}}{\sqrt{2}}+e^{i(\phi_{NL,2}^{'}+\phi_L)} e^{-\gamma/2}  X \right)
\end{equation}
where $X=\frac{E_1+iE_4 }{2}$. Without any loss of generality, we can set $\phi_L=0$ and take $\phi_i=\phi_{NL,i}$ and similarly, $\phi_i^{'}=\phi_{NL,i}^{'}$. To express Eq. 6 in the form of a discrete-time iterative equation, we need to declare the parameter ‘$t_R$’ which characterizes the total time the electromagnetic field takes to complete one complete round trip across the attenuated feedback loop. Taking time steps equal to $t_R$, we can rewrite the evolution of field amplitude $E_1$ described by Eq. 6 in the form of an iterative equation as given below.
\begin{equation}
E_{1,j+1}=e^{i\phi_{1,j}} e^{-\gamma/2} \left(\frac{iE_{in}}{\sqrt{2}}+e^{i\phi_{2,j}^{'}} e^{-\gamma/2}  X_{j}\right)
\end{equation}
where $X_{j}=\frac{E_{1,j}+iE_{4,j}}{2}$. This equation relates the electric field amplitude $E_1$  in the $(j+1)$-th iteration with the same in the $j$-th iteration. Similarly, we can express the field amplitude $E_4$ in the form of a discrete-time iterative equation as follows
\begin{equation}
E_{4,j+1}=e^{i\phi_{4,j}} e^{\gamma/2} \left(\frac{iE^{'}_{in}}{\sqrt{2}}+e^{i\phi_{3,j}^{'}} e^{\gamma/2}  Y_{j}\right)
\end{equation}
where $Y_{j}=\frac{iE_{1,j}+E_{4,j} }{2}$. One important point that must be noted here is that the time-delayed feedback loops in our system are driven periodically by the field amplitudes $E_{in}$ and $E^{'}_{in}$. The time period between each drive is the time taken by the field amplitudes to complete one round trip across the feedback loop $t_R$. Hence, we can say that the field amplitudes $E_{in}$  and $E^{'}_{2}$ must reach the coupler region ‘\textbf{\textit{DC I}}’ at the same instance. Similarly, $E^{'}_{in}$  and $E^{'}_{4}$ must reach the coupler region ‘\textbf{\textit{DC III}}’ at the same instance.
\setcounter{secnumdepth}{0}
\section{REFERENCES}
\begin{enumerate} [label={[\arabic*]}]
	\item K. Ikeda, Opt. Commun. 30, 257 (1979).
	\item K. Ikeda, H. Daido and O. Akimoto, Phys. Rev. Lett. 45, 709 (1980).
	\item H. Haken, Phys. Lett. A 53, 77 (1975).
	\item G. H. M. van Tartwijk  and G. P. Agrawal, Opt. Commun. 133, 565 (1997).
	\item R. Lang and K. Kobayashi, 1EEE J. Quantum Electron. 16, 347 (1980).
	\item A. Uchida, \textit{Optical Communication with Chaotic Lasers- Applications of Nonlinear Dynamics and Synchronization}, Wiley, New York (2012).
	\item J. Mørk, B. Tromborg, and J. Mark, IEEE J. Quantum Elec. 28, 93 (1992).
	\item G. D. Van Wiggeren and R. Roy, Science 279, 1198 (1998).
	\item I. Fischer, Y. Liu, P. Davis, Phys. Rev. A 62, 011801 (2000).
	\item B. Haegeman \textit{et al.}, Phys. Rev. E 66, 046216 (2002).
	\item J. R. Terry \textit{et al.}, Phys. Rev. E 59, 4036  (1999).
	\item G. Vemuri and R. Roy, Phys. Rev. A 39, 4668 (1989).
	\item B. McNamara, K. Wiesenfeld and R. Roy, Phys. Rev. Lett. 60, 2626 (1988).
	\item J. Zamora-Munt, C. Masoller, J. Garcia-Ojalvo and R. Roy, Phys. Rev. Lett. 105, 264101 (2010).
	\item F. Duport, B. Schneider, A. Smerieri, M. Haelterman, and S. Massar, Opt. Exp. 20, 22783 (2012).
	\item J. D. Hart, L. Larger, Murphy Thomas E. and R. Roy, Phil. Trans. R. Soc. A. 377, 20180123 (2019).
	\item C. Sugano, K. Kanno, and A. Uchida, IEEE J. Sel. Top. Quantum Electron. 26, 1500409 (2020).
	\item S. Sunada and A. Uchida, Sci. Rep. 9, 19078 (2019).
	\item K. Takano \textit{et al} , Opt. Exp. 26, 13521 (2018).
	\item C. M. Bender, S. Boettcher, Phys. Rev. Lett. 80, 5243 (1998).
	\item C. M. Bender, S. Boettcher, P. N. Meisinger, J. Math. Phys. 40, 2201 (1999).
	\item C. M. Bender, D.C. Brody, H. F. Jones, Phys. Rev. Lett. 89, 270401 (2002).
	\item C. M. Bender, Rep. Prog. Phys 70, 947 (2007).
	\item R. El-Ganainy, K. G. Makris, D. N. Christodoulides and Z. H. Musslimani, Opt. Lett. 32, 2632 (2007).
	\item C. E. R\"uter \textit{et al.}, Nat. Phys. 6, 192 (2010).
	\item Z. Lin \textit{et al.}, Phys. Rev. Lett. 106, 213901 (2011).
	\item X. Xu, Y. Liu, C. Sun, and Y. Li, Phys. Rev. A 92, 013852 (2015).
	\item X. L\"u, H. Jing, J. Ma, and Y. Wu, Phys. Rev. Lett. 114, 253601 (2015).
	\item M. -A. Miri, A. Regensburger, U. Peschel and D. N. Christodoulides, Phys. Rev. A 86, 023807 (2012).
	\item J. Ren \textit{et al.}, Opt. Express 26, 27153 (2018).
	\item A. K. Sarma, M.-A. Miri, Z. H. Musslimani, D. N. Christodoulides, Phys. Rev. E 89, 052918 (2014).
	\item M. -A. Miri \textit{et al.}, Phys. Rev. A 86, 033801 (2012).
	\item A. Govindarajan, A. K. Sarma and M. Lakshmanan, Opt. Lett. 44, 663  (2019).	
	\item S. Assawaworrarit, X. Yu and S. Fan, Nat. 546, 387 (2017).
	\item M. Sarisaman, Phys. Rev. A 95, 013806 (2017).
	\item J. P. Deka and A. K. Sarma, Appl. Opt. 57, 1119 (2018).
	\item Y. Takasu, T. Yagami, Y. Ashida, R. Hamazaki, Y. Kuno and Y. Takahashi, Prog. Theor. Exp. Phys. 2020, 12 (2020).
	\item J. P. Deka, A. K. Sarma, A. Govindarajan, M. Kulkarni, Nonlinear Dyn 100, 1629 (2020).	
	\item J. P. Deka, A. K. Sarma, Nonlinear Dyn 96, 565 (2019).
	\item D. R. Solli, C. Ropers, P. Koonath, and B. Jalali, Nat. 450, 1054 (2007).
	\item K. Hammani, C. Finot, J. M. Dudley, and G. Millot, Opt. Express 16, 16467 (2008).
	\item F. T. Arecchi, U. Bortolozzo, A. Montina, and S. Residori, Phys. Rev. Lett. 106, 153901 (2011).
	\item M. S. Ruderman, Eur. Phys. J. Special Topics 185, 57 (2010).
	\item C. Bonatto \textit{et al.}, Phys. Rev. Lett. 107, 053901 (2011).
	\item S. L. Kingston, K. Thamilmaran, P. Pal, U. Feudel, S. K. Dana, Phys. Rev. E 96, 052204 (2017).
\end{enumerate}
\end{document}